\RequirePackage{fix-cm}
\documentclass[smallextended]{svjour3}       
\usepackage[italian,english]{babel}
\smartqed  
\usepackage{amsmath}
\usepackage{amssymb}
\usepackage{lipsum}
\usepackage{dsfont}
\usepackage{multirow}
\usepackage{afterpage}
\usepackage{amsfonts}
\usepackage{bm}
\usepackage{babel}
\usepackage{hyperref}
\usepackage{graphicx}
\usepackage{color}
\usepackage{graphicx}
\begin{document}

\title{Weak Measurement effects on dynamics of quantum correlations in a Two-atom System in Thermal Reservoirs
}


\author{N. Ananth \and R. Muthuganesan   
\and  V. K. Chandrasekar }


\institute{PG \& Research Department of Physics,  Bishop Heber College (Affiliated to Bharathidasan University), 
Tiruchirappalli-620 017, Tamil Nadu, India\at
             \email{ananth.ph@bhc.edu.in}           \and 
Center for Nonlinear Science \& Engineering, School of Electrical \& Electronics Engineering,
SASTRA Deemed University, Thanjavur, Tamil Nadu 613 401, India\at
              Tel.: +91-8760372825\\
              \email{rajendramuthu@gmail.com}           
           \and
         Center for Nonlinear Science \& Engineering, School of Electrical \& Electronics Engineering,
SASTRA Deemed University, Thanjavur, Tamil Nadu 613 401, India\at
              \email{chandru@gmail.com}
}

\date{Received: date / Accepted: date}

\maketitle

\begin{abstract}
The dynamical behavior of quantum correlations captured by different forms Measurement-Induced Nonlocality (MIN) between two atoms coupled with thermal reservoirs is investigated and compared with the entanglement. It is shown that the MIN quantities are more robust, while noise cause sudden death in entanglement. Further, we quantified the quantum correlation with weak measurement and the effect of measurement strength is observed. The role of mean photon number and weak measurement on quantum correlation is also highlighted. 
\end{abstract}

\keywords{Entanglement \and Quantum correlation  \and Dynamics \and Weak measurement \and Projective measurements \and Nonlocality}

\section{Introduction}
\label{intro}
Generation and protection of quantum correlation in various quantum systems is 
one of the fundamental areas of interest in quantum communication and information theory \cite{Einstein,Schrodinger,Nielsen,Knill,Datta2005,Datta2005PRL,Henderson,bennett2000}. 
In particular, make use of such a correlation in open systems is a cumbersome task due to decoherence processes \cite{breuer2000}. This  inevitable interaction between a quantum system and its environment would destroy quantum correlations between the systems  and that reduce the quantum advantages of the system. Further quantum correlation dynamics provide the knowledge to  design suitable mechanisms to protect quantum correlations during information processing \cite{zurek2003}.  Recently, researchers paid more attention to identifying different types of quantum correlations in various quantum systems  \cite{ollivier2001,luo2010,luo2008,luo2011,xiong2014}. In particular, generation of nonlocal  correlation beyond the entanglement,  because entanglement does not account for all of the non-classical properties of quantum correlations.   

The present work will focus on the dynamics of quantum correlations beyond the entanglement in open quantum systems. Yi and Sun made the first  attempt to study the dynamics of entanglement in open systems \cite{yi1999}. Subsequently,  the dynamics of entanglement by considering two coupled and initially entangled harmonic oscillators in environment is investigated \cite{rajagopal2001}. Zyczkowski et al. studied dynamics of entanglement for the bipartite systems under the action of decaying channel \cite{zyczko2001}. Further, it is shown that entanglement might disappear at finite times, while coherence vanish asymptotically in the same environment  \cite{rajagopal2001,zyczko2001}. 

It is worth mentioning that weak measurement allowing us to probe into the quantum system with minimum disturbance, which was introduced by Aharonov etal. \cite{ahar1988}. A quantum measurement with any number of outcomes can be contructed as a sequence of weak measurements \cite{oresh2005}.  Binding the concept of dynamics of quantum correlation and weak measurement, there were several results  reported as follows.  
 
 The dynamics of quantum correlation for two isolated atoms in their own thermal reservoir are studied by using quantum discord with weak measurement \cite{bai2020}. It is shown that the quantum correlation quantified  by a sequence of geometric discord with weak measurements  \cite{li2016}. Further, the comparison between dynamics of quantum discord and entanglement of two-atom in thermal reservoir is also investigated \cite{yan2014}.  However, from a computational point of view, the evaluation of quantum discord is a tough  task due to the complicated optimization procedure \cite{modi2012}. Consequently, a lot of alternative measures of quantum correlation have been proposed, such as geometric quantum discord, quantum deficit,  measurement induced disturbance (MID), measurement induced nonlocality (MIN) \cite{ollivier2001,luo2010,luo2011}. These measures are also quantum measurement dependent. Further, it is also shown that MIN is more robust than the concurrence to decoherence \cite{ma2016}.  Despite the correlation dynamics has been broadly studied in open quantum systems, the problem is still open for the effect of the environment on the different forms of MIN. Therefore MIN is thus taken as dynamical quantity with and without weak measurements,  and we study the evolution of MINs under the influence of environment. 

The article is organized as follows.  In Sec. \ref{correlation}, we define the quantifiers of quantum correlation studied in this paper. In Sec. \ref{model}, we introduce the theoretical model under our investigation and the notion of intrinsic decoherence in a quantum system. The dynamics of entanglement and MINs presented in Sec. \ref{dynamics}. Finally, the conclusions are given in Sec. \ref{Concl}.

\section{Quantum correlation measures}

\label{correlation}

\noindent \textit{Entanglement}: Concurrence has been recognized as a powerful measure of entanglement for arbitrary two-qubit states \cite{Wootters}. 
It can be defined for a bipartite composite state 
$\rho$ shared by the subsystem $a$ and $b$ in the Hilbert space $\mathcal{H}=\mathcal{H}^a\otimes\mathcal{H}^b$, that is 
\begin{equation}
C(\rho )=\text{max}\{0,~\lambda_1-\lambda_2-\lambda_3-\lambda_4\},
\end{equation}
where $\lambda_i$ are the square root of eigenvalues of matrix 
$R=\rho(\sigma_y \otimes \sigma_y) \rho* (\sigma_y \otimes \sigma_y)$ 
arranged in decreasing order. The symbol $*$ denotes the usual complex conjugate in the computational basis. The concurrence varies from $0$ to $1$ with minimum and maximum values correspond to separable and maximally entangled states respectively.\\

\noindent \textit{Measurement induced nonlocality}:
Next, we employ measurement-induced nonlocality as a second  measure of quantum correlations (QC). MIN, captures the maximal nonlocal effects of bipartite state due to locally invariant projective measurements,  is defined as the maximal distance between the quantum state of our consideration and the corresponding state after performing a local measurement on one of the subsystems i.e., \cite{luo2011}
\begin{equation}
 N_2(\rho ) =~^{\text{max}}_{\Pi ^{a}}\| \rho - \Pi ^{a}(\rho )\| ^{2},
\end{equation}
where $\|\mathcal{O}\| ^{2}=\text{Tr}(\mathcal{O}\mathcal{O}^{\dagger})$ is Hilbert-Schmidt norm of operator $\mathcal{O}$ and the maximum is taken over the von Neumann projective measurements on subsystem $a$ which does not change the state.  Mathematically it is defined as $\Pi^{a}(\rho) = \sum _{k} (\Pi ^{a}_{k} \otimes   \mathds{1} ^{b}) \rho (\Pi ^{a}_{k} \otimes    \mathds{1}^{b} )$, with $\Pi ^{a}= \{\Pi ^{a}_{k}\}= \{|k\rangle \langle k|\}$ being the projective measurements on the subsystem $a$, which do not change the marginal state $\rho^{a}$ locally i.e., $\Pi ^{a}(\rho^{a})=\rho ^{a}$. If $\rho^{a}$ is non-degenerate, then the maximization is not required.   

In general, the arbitrary two qubit state in the Bloch representation can be written as 
\begin{equation}
\rho=\frac{1}{2}\left[ X_0 \otimes Y_0+\sum_{i=1}^3 x_i (X_i\otimes Y_0)+\sum_{j=1}^3 y_j (X_0 \otimes Y_j)+\sum_{i,j=1}^3 t_{ij} X_i \otimes Y_j \right], 
\label{Equations} 
\end{equation}
where $x_i=\text{Tr}(\rho(X_i\otimes Y_0))$, $y_j=\text{Tr}(\rho(X_0 \otimes Y_j))$ are the components of Bloch vector and $t_{ij}=\text{Tr}(\rho(X_i \otimes Y_j))$ being real matrix elements of correlation matrix $T$. In bipartite state space, the orthonormal operators in respective state spaces are $\{X_0,X_1,X_2,X_3 \}=\{\mathds{1},\sigma_1,\sigma_2,\sigma_3 \}/\sqrt{2} $ and $\{Y_0,Y_1,Y_2,Y_3 \}=\{\mathds{1},\sigma_1,\sigma_2,\sigma_3 \}/\sqrt{2} $, where $\sigma_i$ are the Pauli matrices. MIN has a closed formula as 
\begin{equation}
N_2(\rho ) =
\begin{cases}
\text{Tr}(TT^t)-\frac{1}{\| \textbf{x}\| ^2}\textbf{x}^tTT^t\textbf{x}& 
 \text{if} \quad \textbf{x}\neq 0,\\
 \text{Tr}(TT^t)- \lambda_{\text{min}}&  \text{if} \quad \textbf{x}=0,
\end{cases}
\label{HSMIN}
\end{equation}
where $\lambda_{\text{min}}$ is the least eigenvalue of matrix $TT^t$, the superscripts $t$ stands for the transpose and the vector $\textbf{x}=(x_1,x_2,x_3)^t$.\\ 

\noindent \textit{Trace distance-based MIN}:
It is a well-known fact that the MIN based on the Hilbert-Schmidt norm is not a credible measure in capturing nonlocal attributes of a quantum state due to the local ancilla problem \cite{Chang2013}. A  natural way to circumvent this issue is defining MIN in terms of contractive distance measure. Another alternate form of MIN is based on trace distance \cite{HUTMIN},  namely, trace MIN (T-MIN)  which resolves the local ancilla problem  \cite{Piani2012}. 
It is defined as
\begin{equation}
N_1(\rho):= ~^{\text{max}}_{\Pi^a}\Vert\rho-\Pi^a(\rho)\Vert_1,
\end{equation} 
where $\Vert A \Vert_1 = \text{Tr}\sqrt{A^{\dagger}A}$ is the trace norm of operator $A$. Here also, the maximum is taken over all von Neumann projective measurements. Without loss of generality, we can rewrite  Eq. (\ref{Equations}) as 
\begin{equation}
\rho=\frac{1}{4}\left[ \mathds{1} \otimes \mathds{1}+\sum_{i=1}^3 x_i (\sigma_i\otimes \mathds{1})+\sum_{j=1}^3 y_j (\mathds{1} \otimes \sigma_j)+\sum_{i,j=1}^3 c_{ij} \sigma_i \otimes \sigma_j\right], \label{EQ} 
\end{equation}
where $c_{ij}=\text{Tr}(\rho(\sigma_i\otimes \sigma_j))$.   For the above system, the closed formula of trace MIN $N_1(\rho)$ is given as 
\begin{equation}
N_1(\rho)=
\begin{cases}
\frac{\sqrt{\chi_+}~+~\sqrt{\chi_-}}{2 \Vert \textbf{x} \Vert_1} & 
 \text{if} \quad \textbf{x}\neq 0,\\
\text{max} \lbrace \vert c_1\vert,\vert c_2\vert,\vert c_3\vert\rbrace &  \text{if} \quad \textbf{x}=0,
\end{cases}
\label{TMIN}
\end{equation}
where $\chi_\pm~=~ \alpha \pm 2 \sqrt{\tilde{\beta}}~\Vert \textbf{x} \Vert_1 ,\alpha =\Vert \textbf{c} \Vert^2_1 ~\Vert \textbf{x} \Vert^2_1-\sum_i c^2_i x^2_i,\tilde{\beta}=\sum_{\langle ijk \rangle} x^2_ic^2_jc^2_k,~ \vert c_i \vert $ are the absolute values of $c_i$ and the summation runs over cyclic permutation of 
$\lbrace 1,2,3 \rbrace$.

Next, we briefly review about the quantum correlation from the perspectives of weak measurements. The weak measurement operators are given by \cite{Singh}
\begin{align}
P(+)=t_1\Pi_1+t_2\Pi_2, ~~~~~~P(-)=t_2\Pi_1+t_1\Pi_2,
\end{align}
where 
\begin{align}
t_1=\sqrt{\frac{1-\text{tanh}x}{2}}~~~~~ \text{and}~~~~~~t_2=\sqrt{\frac{1+\text{tanh}x}{2}}
\end{align}
with $x$ denoting the strength in the weak measurement process, $\Pi_1$ and $\Pi_2$ are two orthogonal projectors with $\Pi_1+\Pi_2=\mathds{1}$. We can easily show that the weak measurement will reduce to orthogonal projectors when $x\rightarrow \infty $   and $P^{\dagger}(+)P(+)+P^{\dagger}(-)P(-)=\mathds{1}$. The post-measurement state $\Omega(\rho)$ after the weak measurement is 
\begin{align}
\Omega(\rho)=P(+)\rho P(+)+P(-)\rho P(-). 
\end{align}
The MIN is based on the weak measurement is defined as 
\begin{equation}
N_p(\rho):= ~^{\text{max}}_{\Omega}\Vert\rho-\Omega(\rho)\Vert_p^p.
\end{equation} 

The actual MIN is related MIN based on the weak measurement is  \cite{Li} 
\begin{align}
N_{2W}(\rho)=(1-t_1t_2)N_{2} (\rho).
\end{align}
Similarly, the trace MIN based on the weak measurement is 
\begin{align}
N_{1W}(\rho)=(1-t_1t_2)N_{1} (\rho).
\end{align}
Next, we show that the dynamical behaviour of correlation measure can be extracted by the MIN with weak measurements. 

\section{The Model and Solution}

\label{model}

To investigate the dynamics of standard quantum correlation measure and quantum correlation with weak measurements, we consider a dissipative model, namely,  two two-level atoms that are each trapped separately in one of the two spatially separated thermal reservoir. Using the standard quantum reservoir theory, time evolution of the density operator for the system after tracing out the reservoir obeys the following master equation 
\begin{align}
 \dot{\rho}=&\sum_{i=1,2} \bigg\{-\frac{1}{2}\gamma_i(n_i+1)\left[\sigma_+^i\sigma_-^i\rho-2\sigma_-^i\rho\sigma_+^i+\rho\sigma_+^i\sigma_-^i \right]\notag \\
&-\frac{1}{2}\gamma_in_i\left[\sigma_+^i\sigma_-^i\rho-2\sigma_-^i\rho\sigma_+^i+\rho\sigma_+^i\sigma_-^i \right] \bigg\} , 
\end{align}
where $\gamma_i$ is the spontaneous emission rate of atom $i$, $n_i$ are mean thermal photon numbers of their local thermal reservoirs separately and $\sigma_+^i (\sigma_-^i)$ is the usual raising (lowering) operator of atom $i$. Since the dynamics of a quantum system depends on several factors such as the  initial states of the system and environment, the types of interaction between system and environment, and the structure of the reservoir.

Let us consider a mixture of the Bell state and an excited state as initial state to show the evolution of entanglement and MINs, that is  
\begin{align}
 \rho(0)=(1-r)|11 \rangle \langle 11| + r |\Phi^+ \rangle \langle \Phi^+|,
\end{align}
where $|\Phi^+ \rangle=\frac{1}{\sqrt{2}}(|01 \rangle +|10 \rangle $ is maximally entangled state and $0\leq r\leq 1$ determines the degree of purity of initial state. When $r=0$, the initial state is $\rho(0)=|11 \rangle \langle 11| $ a separable state and does not have any quantum advantage  and $r=1$ the initial  state is pure and maximally entangled state. For simplicity, we set $n_1=n_2=n$ and $\gamma_1=\gamma_2=\gamma$. Then the solution of the master equation for the given initial condition is
\begin{align}
\rho(t) = 
\begin{pmatrix}
\rho_{11}(t) & 0 & 0 & 0\\
0 & \rho_{22}(t) & \rho_{23}(t) & 0\\
0 & \rho_{32}(t) & \rho_{33}(t) & 0\\
0 & 0 & 0 & \rho_{44}(t) 
\end{pmatrix},
\end{align}
where 
\begin{align}
\rho_{11}(t)=&\frac{1}{2(2n+1)^2}\{2n^2-2n\left[r(2n+1)-2(n+1)\right]\mathrm{e}^{-(2n+1)\gamma t} \notag\\&-\left[r(n+1)(4n+2)-2(n+1)^2\right]\mathrm{e}^{-2(2n+1)\gamma t} \},  \nonumber \\
\rho_{22}(t)=&\rho_{33}(t)=\frac{1}{2(2n+1)^2}\{2n(n+1)-[r(2n+1)-2(n+1)]\mathrm{e}^{-(2n+1)\gamma t}\notag\\& +\left[r(n+1)(4n+2)-2(n+1)^2\right]\mathrm{e}^{-2(2n+1)\gamma t}\}, \nonumber \\
\rho_{44}(t)=&\frac{1}{2(2n+1)^2} \{2(n+1)^2+2(n+1)[r(2n+1)-2(n+1)]\mathrm{e}^{-(2n+1)\gamma t}\notag\\&-[r(n+1)(4n+1)-2(n+1)^2]\mathrm{e}^{-2(2n+1)\gamma t}\},  \nonumber \\
\rho_{23}(t)=&\rho_{32}(t)=\frac{r}{2}\mathrm{e}^{-(2n+1)\gamma t}.  \nonumber
\end{align}
 The entanglement and MINs of the time evolved state  $\rho(t)$ is computed as 
\begin{align}
C(\rho)=2 ~\text{max} \{ 0, | \rho_{23}|-\sqrt{\rho_{11} \rho_{44}} \},  ~~~~~~~~~~~~~~~~~~~~~\\
N_2(\rho)=2 ~| \rho_{23}|^2 ~~~~~~~~~ \text{and}~~~~~~~~~~~ N_1(\rho)=2 ~| \rho_{23}|.
\label{elements}
\end{align}
Substituting the matrix elements in the above equation, we obtain the entanglement and MIN of the time evolved state.

\section{Quantum correlation dynamics of the two atom systems}
\label{dynamics}
\begin{figure}[!ht]
\centering 
\includegraphics[width=0.5\linewidth]{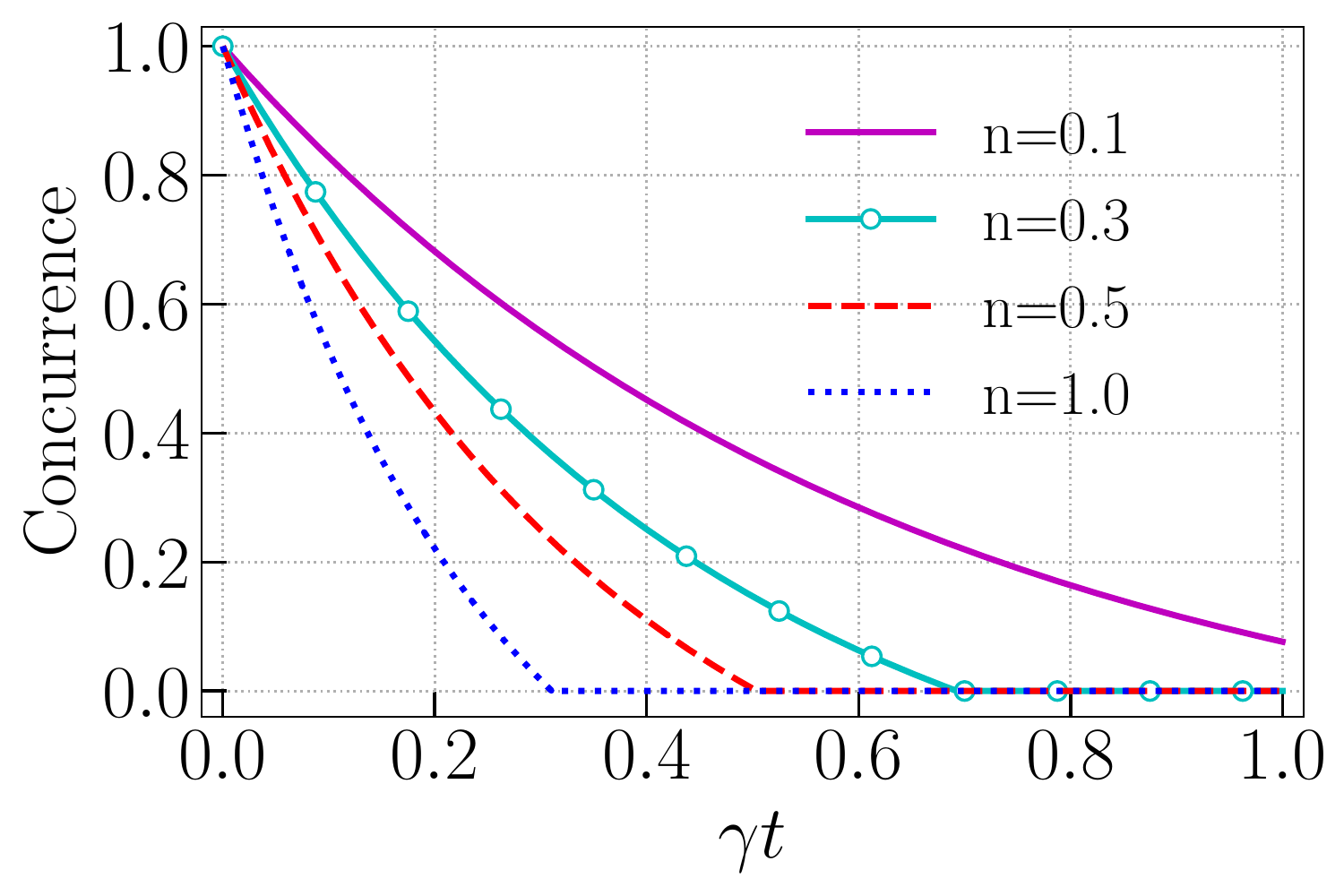}
\centering  \includegraphics[width=0.5\linewidth]{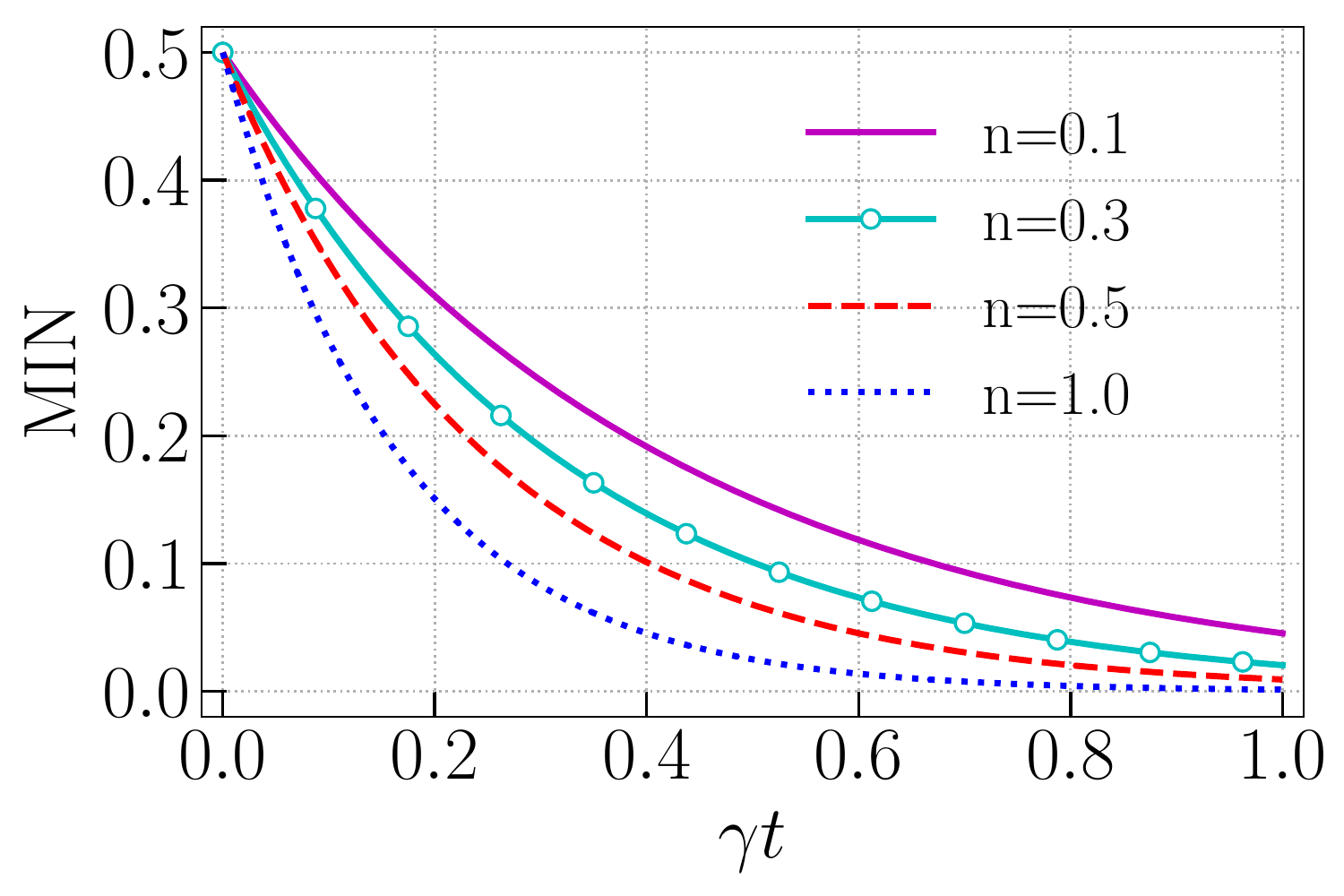}
\centering \includegraphics[width=0.5\linewidth]{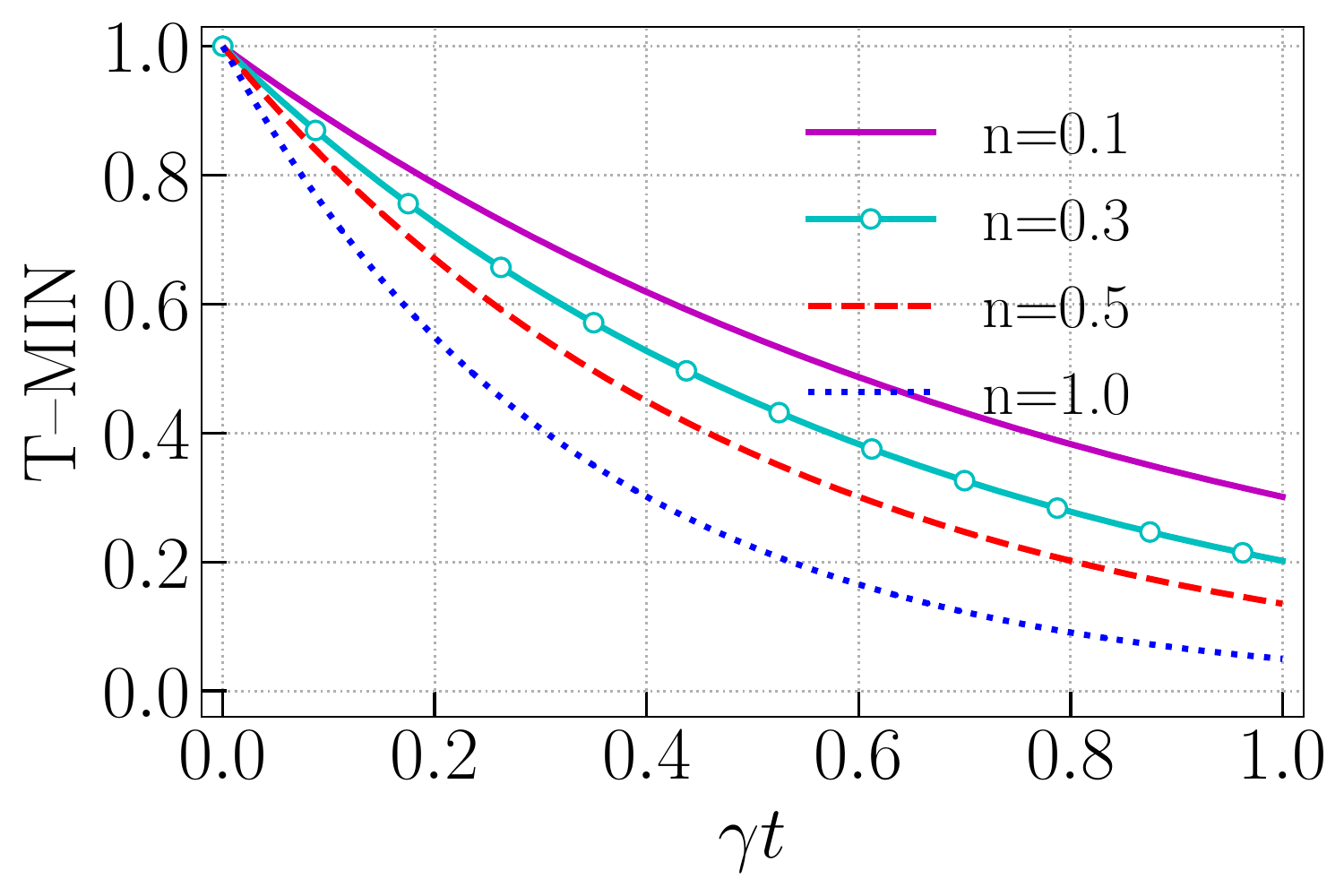}
\caption{(color online) Entanglement and MINs of 
$\rho(t)$ for the different mean photon numbers 
$n$ with $r=1$.}
\label{fig123}
\end{figure}

To study the dynamics of quantum correlations,  we plot the entanglement measured by concurrence and the quantum correlation beyond entanglement quantified by different forms of measurement-induced nonlocality (MIN) as a function of scaled time $\gamma t$ for the fixed mean thermal photon number in Fig. \ref{fig123}. 

First, we set $r=1$, then the initial state is pure and maximally entangled state. For $n=1$, we observe that at time $t=0$, the entanglement is maximum, as time increases the entanglement decreases monotonically and vanishes at a critical point. After the critical point, the entanglement between the qubits remains zero and this is known as sudden death of entanglement. On the other hand, the companion quantities MIN and T-MIN are also maximum at $t=0$, and further increase of time cause monotonic decrease in the quantum correlations. In contrast to the entanglement, the MIN is vanishing only at the asymptotic limit and the trace MIN is non-zero even in the asymptotic limit $t  \rightarrow \infty$. This is due to the fact that  the quantum system interacting with their dissipation environments. The above observation also highlights the robustness of MIN quantities. Further, it also emphasizes that an efficient quantum  information processing based on the MIN offers more resistance to an external perturbation.  

To assert the prominent role of mean photon number, we plotted the entanglement and MINs for the different mean photon numbers such as $n=0.5, 0.3~ \text{and}~ 0.1$. From Fig. \ref{fig123}(a), here also we observe that the  entanglement suffers a sudden death at a certain time point in a thermal reservoir of nonzero mean photon number $n$. Further, we find that the time at which sudden death occurs increasing  with the decrease of mean photon number $n$ and the strength of the entanglement increases with increase of mean photon number.
On the other hand, in order to see how the dissipation affects these two quantum correlations (MIN and trace MIN) differently, we change  the mean photon number in the thermal reservoirs. The decrease in mean photon number greatly enhances the strength of quantum correlation between the subsystems as shown in 
Fig.\ref{fig123}(b) and (c). 
\begin{figure*}[!ht]
\centering\includegraphics[width=0.5\linewidth]{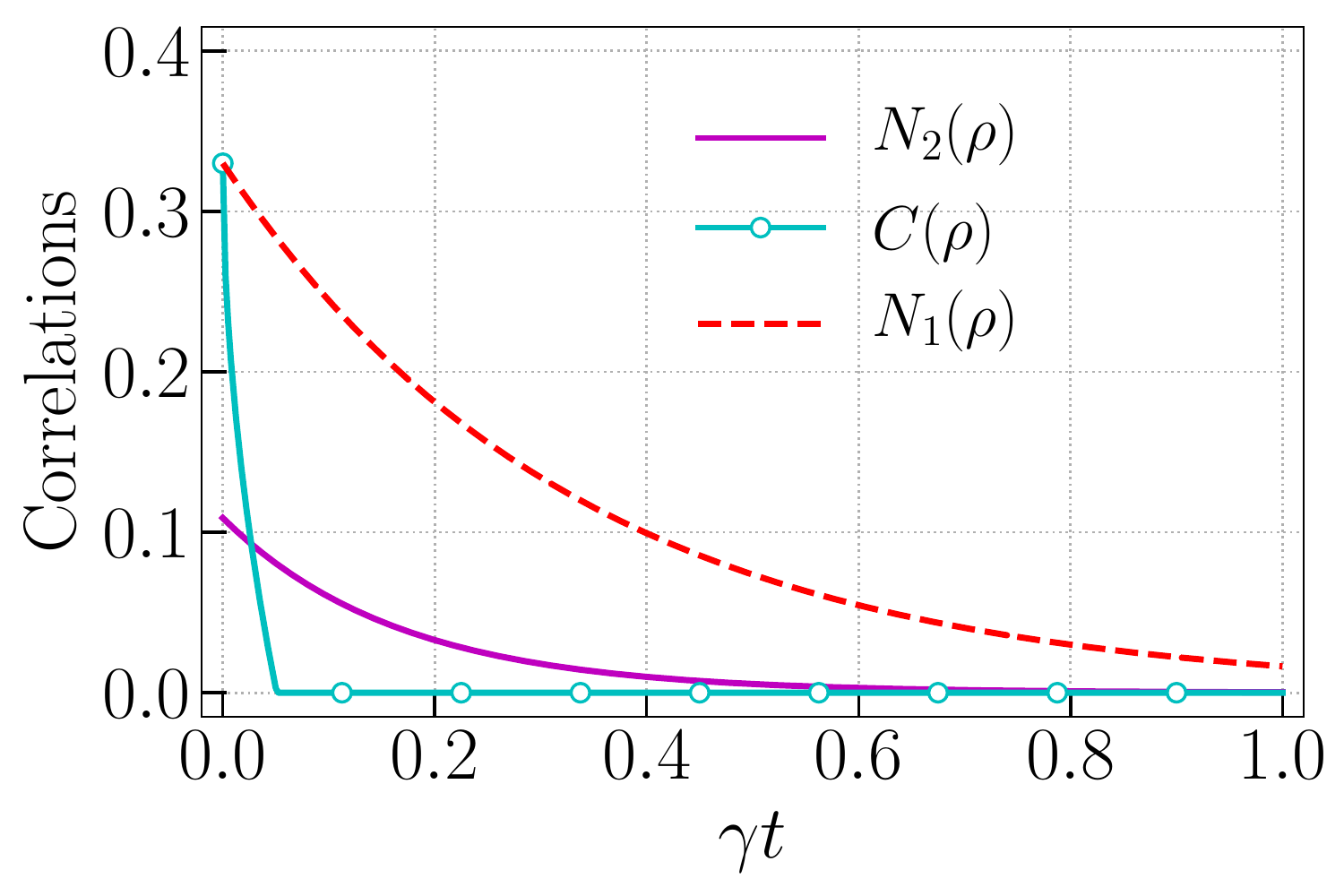}
\centering\includegraphics[width=0.5\linewidth]{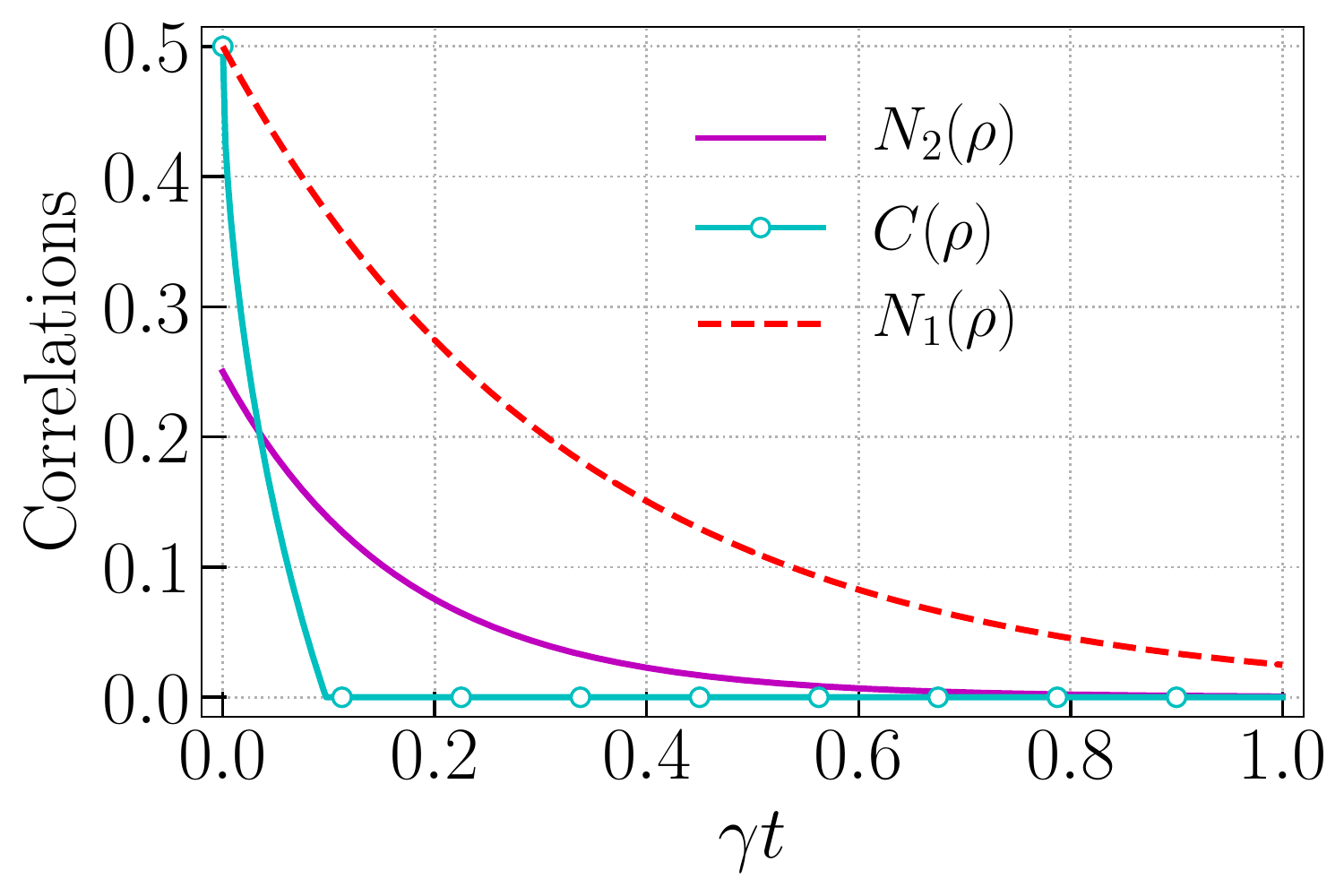}
\centering\includegraphics[width=0.5\linewidth]{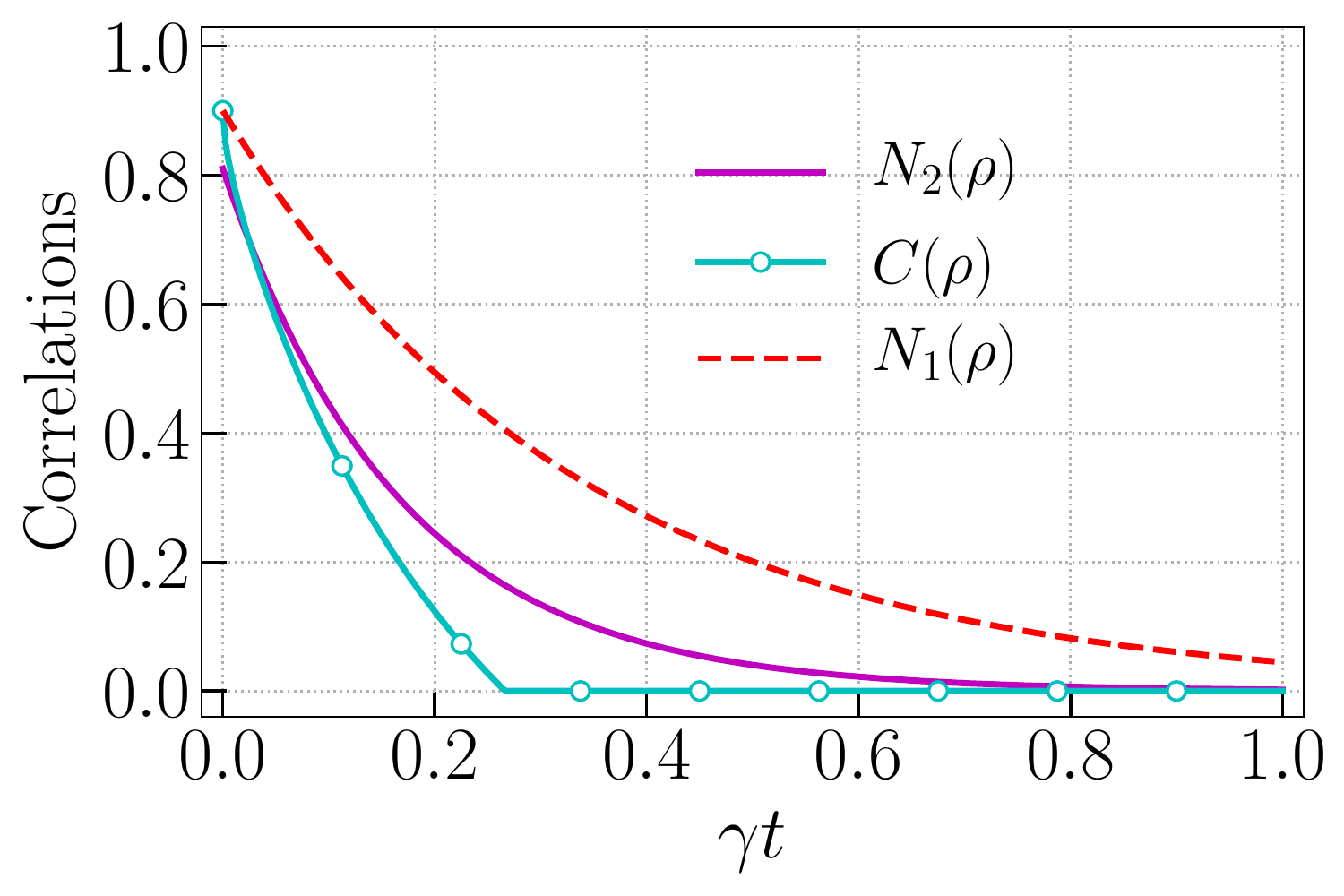}
\caption{(color online) Entanglement and MINs of 
$\rho(t)$ for (Top) $r=0.3$, (middle) $r=0.5$ and   $r=0.9$ (bottom).}
\label{fig456}
\end{figure*}

To understand the effects of purity or mixedness on quantum correlations, in Fig. (\ref{fig456}) we have plotted the MIN and entanglement for the different values of $r$ such as $r=0.3,0.5, \text{and}~ 0.9$. For $0<r<1$, the state is partially  entangled state and all the correlation measures show monotonically decreasing behaviors with respect to time.  It is also observed that the evolution of the entangled state under this  channel exhibits entanglement sudden death i.e., influence of quantum noise reduces the entanglement to zero in finite time. As time increases the MIN and T-MIN decrease from maximum values and showing that quantum correlation vanishes asymptotically.

\begin{figure*}[!ht]
\centering\includegraphics[width=0.4\linewidth]{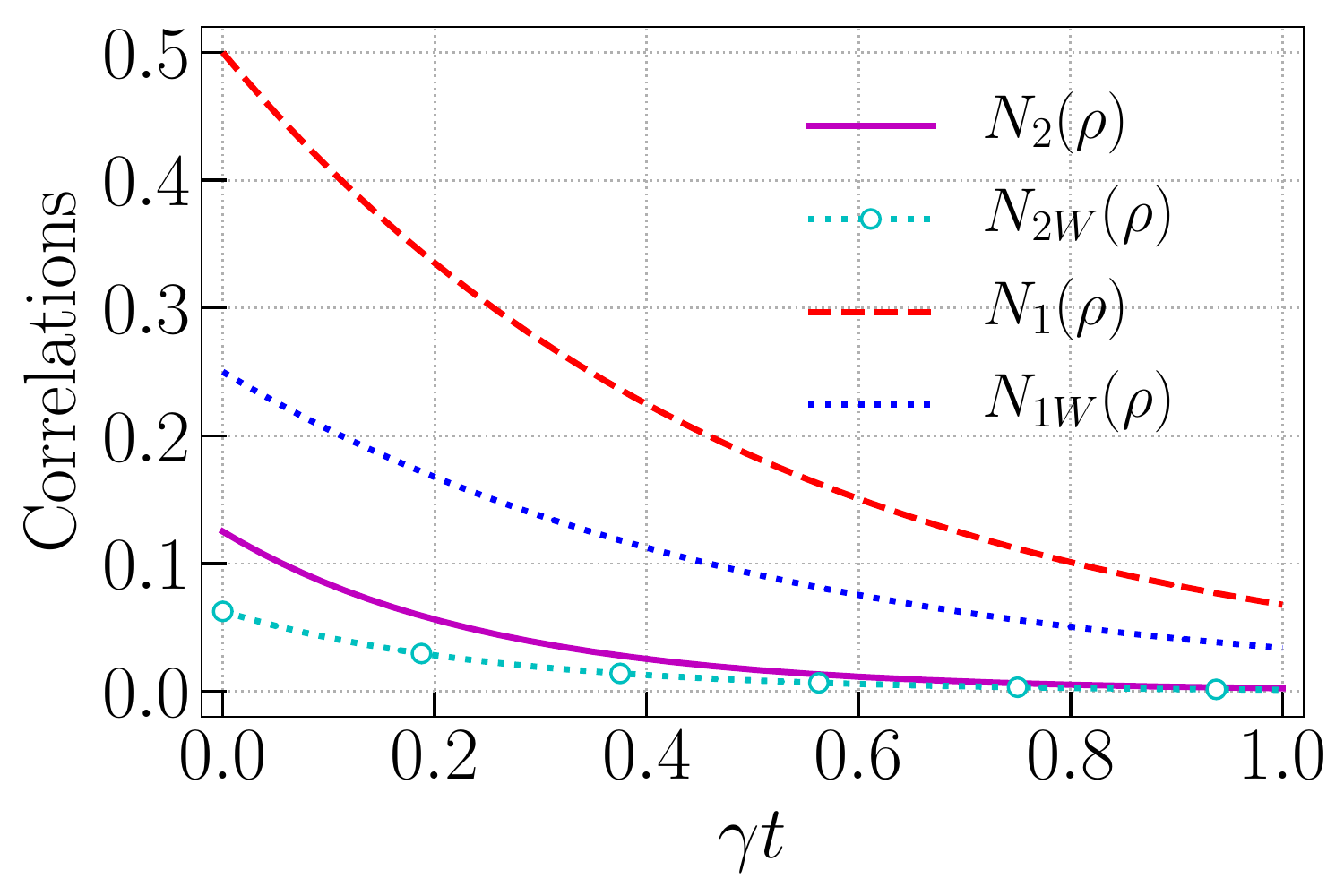}
\centering\includegraphics[width=0.4\linewidth]{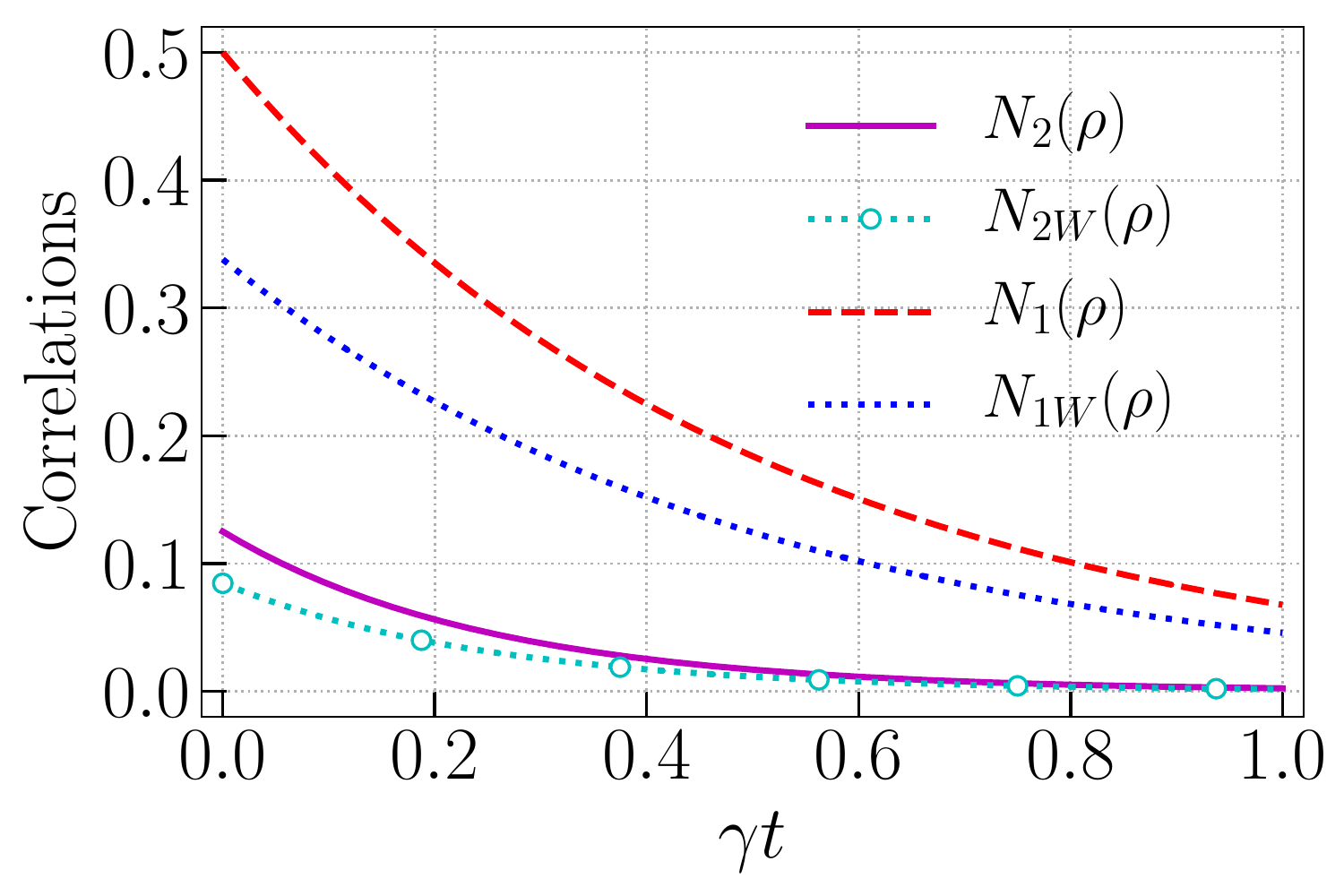}
\centering\includegraphics[width=0.4\linewidth]{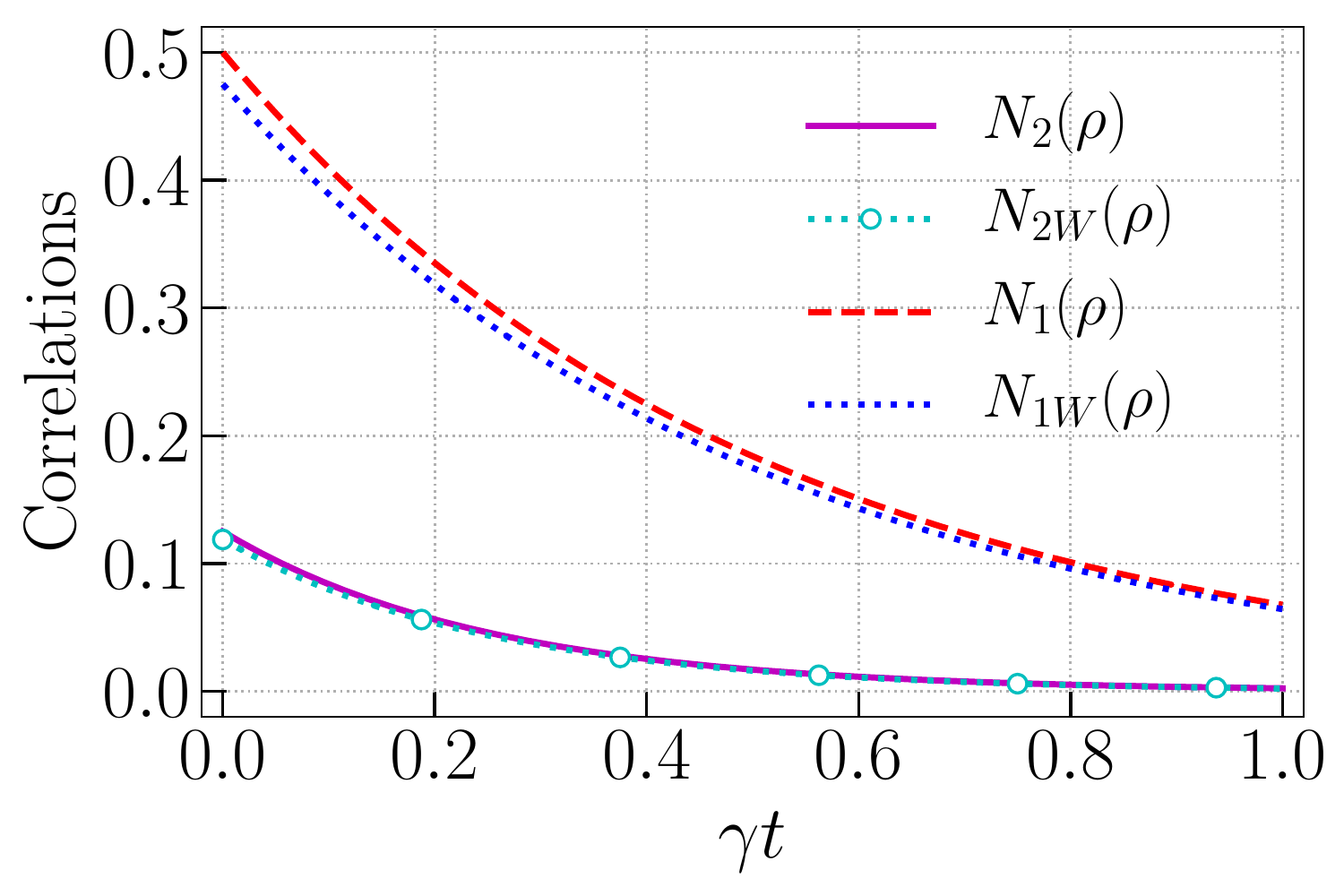}
\centering\includegraphics[width=0.4\linewidth]{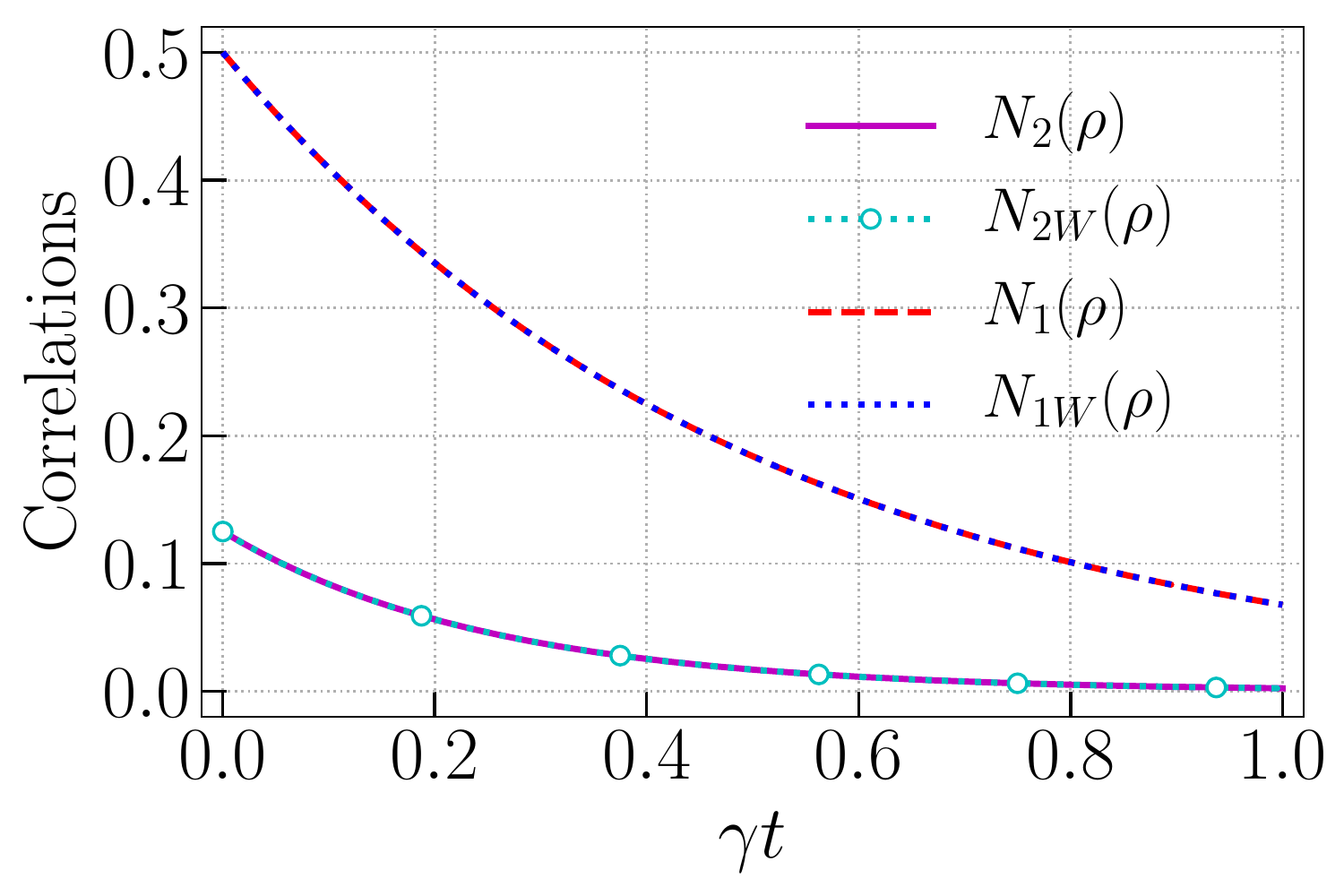}
\caption{(color online) MINs and Weak MINs of  $\rho(t)$ as a function of weak measurement  strength $x$ for the fixed $n=0.5$ and $r=0.5$. Top panel $x=0.1 (Left), x=1 (Right) $ and Bottom panel $x=3 (Left), x=30 (Right)$ }
\label{fig5}
\end{figure*}

In order to understand the effect of weak measurements, we plot the MINs and MINs as a function of weak measurement strength for the fixed values of mean photon numbers $n$ and purity of the quantum state $r$ in Fig. \ref{fig5}. We observe that both MIN and weak MIN exhibit monotonically decreasing behaviors with respect to the scaled time for the fixed $n=r=0.5$. Recently, Singh and Pati introduced super quantum discord (SQD) with weak measurement \cite{Singh} and they have shown that for any state the SQD under weak measurement was always greater than quantum discord revealed by projective measurement. Here, one can be seen clearly for the given values of $x$ that $N_{2W}(\rho)\leq N_{2}(\rho)$ and $N_{1W}(\rho)\leq N_{1}(\rho)$  which is in sharp contrast to the SQD \cite{Singh}. On the other hand, this observation is consistent with the result presented in \cite{Li}, where they proved that weak one-way deficit is smaller than one-way deficit for the given values of $x$. Further, the above results underscore the weak measurements are less invasive than projective measurements, so the Hilbert-Schmidt distance (trace distance) or von Neumann entropy from pre-measurement state to post-weak-measured state is less than that to post-projective-measured state.

While comparing  Fig. \ref{fig5}, we observe that the weak MINs are  monotonically increasing functions of the measurement strength $x$. On the other hand, one can show that SQD is a monotonically decreasing function of the measurement strength $x$. Hence, the above findings reveal that the SQD and weak MIN  capture different nonlocal aspects of quantum system. Further, in the asymptotic limit $x\rightarrow \infty $, the MIN due to the eigenprojective measurements and weak measurement coinciding each other due to the fact that $\Pi^a(\rho)=\Omega(\rho)$. 

\section{Conclusions}
\label{Concl}
In conclusion, we have studied the dynamics of quantum correlations of two qubit systems, each in its own thermal reservoir. Concurrence and measurement induced nonlocality (MIN) are used to quantify the quantum correlations and shown that the correlation measure exhibits monotonic decaying behavior with time. We  observe  that  the entanglement suffered by  sudden  death, while MIN quantities are more robust than  entanglement. This observation implies that the presence  of  nonlocality  (in  terms  of  MIN)  even  in  the  absence  of  entanglement between the local constituents and MINs could capture more quantumness than the entanglement. The quantum correlation quantified with weak measurements and we also found that the MIN and weak MIN tend to decrease as the measurement strength increase. It is shown that the  MIN and weak MIN are identical in measuring the quantum correlation in the asymptotic limit. 

Moreover, we emphasize  that based on our observations, measurement-induced  nonlocality  offer  more  resistance  to the  effect  of  external perturbation  and  MIN is more robust than the entanglement.

%
%

\begin{acknowledgements}

NA acknowledges the financial support received from Bishop Heber College, Tiruchirappalli under Minor Research Grant(MRP/ 1921/ 2019 (BHC)). RM and VKC thank the Council of Scientific and Industrial Research (CSIR), Government of India for the financial support under Grant No. 03(1444)/18/EMR-II.
\end{acknowledgements}



\end{document}